\documentclass[runningheads]{llncs}
\usepackage[T1]{fontenc}
\usepackage{graphicx}
\usepackage{multirow,multicol}
\usepackage{booktabs}
\usepackage{xcolor}
\usepackage{float}
\usepackage{stfloats}
\usepackage{subcaption}

\usepackage[ruled,linesnumbered]{algorithm2e}
\def\BibTeX{{\rm B\kern-.05em{\sc i\kern-.025em b}\kern-.08em
    T\kern-.1667em\lower.ex\hbox{E}\kern-.125emX}}

\begin{document}
\title{EEG-based Mental Imagery Task Adaptation via Ensemble of Weight-Decomposed Low-Rank Adapters}
\titlerunning{EDoRA: Adaptation Method for EEG-based Mental Imagery Tasks}
\author{Taveena Lotey\inst{1\thanks{Corresponding author: Taveena Lotey}} \and
Aman Verma\inst{2} \and Partha Pratim Roy\inst{1}}
\authorrunning{T. Lotey et al.}
\institute{ Department of CSE, , Indian Institute of Technology Roorkee, India \and Department of ME, Indian Institute of Technology Roorkee, India \\
\email{taveena@cs.iitr.ac.in, averma1@me.iitr.ac.in, partha@cs.iitr.ac.in}}
\maketitle              
\begin{abstract}
Electroencephalography (EEG) is widely researched for neural decoding in Brain Computer Interfaces (BCIs) as it is non-invasive, portable, and economical. However, EEG signals suffer from inter- and intra-subject variability, leading to poor performance. Recent technological advancements have led to deep learning (DL) models that have achieved high performance in various fields. However, such large models are compute- and resource-intensive and are a bottleneck for real-time neural decoding. Data distribution shift can be handled with the help of domain adaptation techniques of transfer learning (fine-tuning) and adversarial training that requires model parameter updates according to the target domain. One such recent technique is Parameter-efficient fine-tuning (PEFT), which requires only a small fraction of the total trainable parameters compared to fine-tuning the whole model. Therefore, we explored PEFT methods for adapting EEG-based mental imagery tasks. We considered two mental imagery tasks: speech imagery and motor imagery, as both of these tasks are instrumental in post-stroke neuro-rehabilitation. We proposed a novel ensemble of weight-decomposed low-rank adaptation methods, EDoRA, for parameter-efficient mental imagery task adaptation through EEG signal classification. The performance of the proposed PEFT method is validated on two publicly available datasets, one speech imagery, and the other motor imagery dataset. In extensive experiments and analysis, the proposed method has performed better than full fine-tune and state-of-the-art PEFT methods for mental imagery EEG classification.

\keywords{Electroencephalography \and Deep Learning \and Transfer Learning \and Fine-tuning \and Low-Rank Adaptation.}
\end{abstract}

\section{Introduction}

Brain-computer interfaces (BCIs) represent a developing field of research aimed at establishing direct communication pathways between the brain and external devices. The physiological measure often used for decoding neural signals is electroencephalography (EEG), as it is non-invasive, portable, and economical. There are various experimental BCI paradigms to control external devices through robotic arm \cite{edelman2019noninvasive}, speller system \cite{young2016state}, and exoskeleton \cite{frolov2017post} via particular brain activity in a specific task. Among these paradigms, mental imagery has undergone extensive investigation as a mechanism for controlling BCIs, leveraging the intrinsic brain activity that arises from the voluntary imagination of users \cite{ahn2022multiscale}. Additionally, mental imagery operates independently of external stimuli. Therefore, it supports the development of a user-friendly interface. It reduces fatigue and enhances users' awareness of their surroundings more naturally.

Deep learning (DL) has advanced rapidly in recent years. Many deep learning-based methods have been proposed for enhanced EEG classification. Deep learning-based methods such as Convolutional neural networks (CNN) and Transformers have a large number of parameters, and they require large amounts of data to learn and extract discriminative features. EEG signals have high dimensionality and labor-extensive recording procedures. These challenges result in small datasets that make DL methods prone to overfitting. Additionally, EEG signals suffer from inter- and intra-subject variability \cite{saha2020intra}. One of the possible solutions to reduce inter-subject variability is Domain adaptation (DA). It is defined as using a classifier/model learned on one task with sufficient labeled samples, and this classifier/model adapts to another related task with only a limited amount of training data.

Many works have utilized DA approaches to reduce the distribution shift in two domains of mental imagery tasks. One section of the studies tends to use domain adaptation for intra-subject/cross-session distribution and another for inter-subject/cross-subject distribution among various subjects \cite{lotey2022cross,lee2023selective}. Motor Imagery (MI) is the most researched modality of mental imagery tasks. Moreover, several studies have employed domain adaptation for MI-EEG signal classification. Many recent studies utilized source and target domain correlation assessment approaches to select the source/s similar to target domain \cite{lee2023selective,zhong2023deep,huang2023shallow}. CNN-based feature extractors and classifiers are extensively used in multiple studies on domain adaptation of MI-EEG task classification \cite{miao2023explainable,zhang2022dynamic,liu2022subject}. However, only a few studies explored transformer-based approaches for MI task adaptation \cite{song2023global}. The other modality of mental imagery tasks is speech imagery/imagined speech (SI), which has yet to be explored in domain adaptation.

In the case of inaccessibility of source data and availability of a pre-trained source model, the target model is initialized with parameters of a pre-trained model and subsequently fine-tuned on target domain data \cite{huang2024privacy,xia2022privacy}. With the rapid increase in the number of parameters and depth of the deep neural networks, fine-tuning such large models is a computationally expensive task. This issue is addressed in natural language processing (NLP) and computer vision (CV) domain with parameter-efficient adaptation \cite{houlsby2019parameter}. In parameter-efficient adaptation, new modules are added in between the pre-trained model layers, and these modules are known as adapters. One such adapter method is Low-Rank Adaptation (LoRA) \cite{hu2021lora}. While fine-tuning, only these adapters are trained instead of training the whole pre-trained model on the target domain. Therefore, training only these adapters results in a reduction in the number of trainable parameters. 

To the best of our knowledge, the impact of parameter-efficient adapters is not explored in the EEG classification task. Therefore, our proposed method is based on a high-performance weight-decomposed low-rank adaptation method \cite{liu2024dora}. Additionally, convolutional transformer-based methods have not been explored in mental imagery task classification. Inspired by this, we utilized EEG Conformer, a convolutional transformer method for feature extraction and classification \cite{song2022eeg}. In literature, numerous studies have been proposed to decode the neural activity on a specific mental imagery task \cite{lotey2023feature}. Only a few studies proposed methods to decode neural activity on multiple mental imagery tasks \cite{ahn2022multiscale}. Our method validated the performance of the proposed low-rank adaptation method on two mental imagery datasets, where parameters of a model trained on one dataset are used to fine-tune the model on the other dataset and vice-versa. 

The contributions of this paper are listed as follows:
\begin{itemize}
    \item We demonstrate the parameter efficient fine-tuning (PEFT) based approaches for enhanced mental imagery task classification. To the best of our knowledge, PEFT-based low-rank adaptation (LoRA) is first explored for mental imagery classification tasks in our work. 
    \item We propose EDoRA, a novel ensemble of weight-decomposed low-rank adapters for mental imagery EEG classification tasks for enhanced adaptation performance.
    \item To the best of our knowledge, low-rank adaptation on two categories of mental imagery tasks, i.e., speech and motor imagery, is first explored in our work.
    \item A detailed analysis of parameter-efficient fine-tuning is performed on speech and motor imagery tasks of mental imagery EEG signal. 
\end{itemize}

\section{Related Work}

\subsection{MI Classification}
Several studies have contributed in addressing the problem of inter-subject and intra-subject data distribution variability in motor imagery EEG signal classification. Most of the work in recent years employed source and target domain correlation assessment methods to select the closest source domain for better target domain classification performance \cite{jeon2019domain,zhang2022dynamic,phunruangsakao2022deep,xu2023dual,lee2023selective,zhong2023deep,huang2023shallow}, and convolutional neural network (CNN) based methods are utilized as classifier \cite{zhong2023deep,jeon2019domain,hang2019cross,song2023global,miao2023explainable,zhang2022dynamic,liu2022subject,chen2022single,chen2021multiattention}. Only a few works have studied the impact of transfer learning-based domain adaptation approaches on latest transformer based models \cite{hu2023msatnet}. 

Several studies explored the efficiency of CNN based approaches for domain adaptation in MI-EEG classification. Zhong \textit{et al.} proposed a domain adaptation framework based on correlation alignment of the source and target domain motor imagery EEG data \cite{zhong2023deep}.Then conventional CNN based classifier is used to classify the features in cross-subject settings. Hang et al. proposed a deep domain adaptation network based on CNN with MMD to minimize the source and target distrubution distance, and then applied center-based discriminative feature learning approach to maximize the inter-class distance \cite{hang2019cross}. They jointly optimized the source and target domain data to align the features. Liu  \textit{et al.} proposed a framework for subject adaptation and it includes a CNN based feature extractor, a subject adapter based on MMD to align the source and target domains and reduce the feature distribution shifts \cite{liu2022subject}.

Most of the studies have jointly optimized the source and target domain for MI-EEG classification task. Few studies used transfer learning, i.e., to use model trained on source domain data to optimize the model trained on target domain data. Phunruangsakao \textit{et al.} proposed a deep domain adaptation framework that selects multiple source domains to optimize label classification of single target domain \cite{phunruangsakao2022deep}. This work further experiments by making source parameters inaccessible and making the privacy policy stricter. Inspired by this work, Huang \textit{et al.} proposed a multi-source free domain adaptation framework with attention weighted module for better source and target domain alignment \cite{huang2024privacy}. To keep the privacy of the source domain data, our work uses the parameters of pre-trained source model to fine tune the target model.
 
\vspace{-5mm}
\subsection{SI Classification}
There are few works exploring the classification performance of speech recognition using traditional methods \cite{simistira2023bimodal}. However, the research area of domain adaptation of speech imagery/imagined speech tasks is less explored in the literature. Jimenez \textit{et al.} proposed a deep unsupervised domain adaptation method based on standardization-refinement approach \cite{jimenez2021standardization}. The research area of domain adaptation in EEG-based speech imagery is still required to be explored. With this inspiration, we chose speech imagery as one task of mental imagery EEG classification to validate the effectiveness of our proposed approach.

In literature, only a few works explored the decoding efficiency of transformer based approaches for domain adaptation in MI EEG signal classification \cite{song2023global,hu2023msatnet}. 
Transformer based methods are proved to be efficient than CNN based method in the domain of NLP and CV. With this inspiration, our method used a convolutional transformer based method curated for EEG signal classification with spatial feature extraction power of CNN and temporal feature extraction efficiency of transformers \cite{song2022eeg}. 

Also, to the best of our knowledge, mental imagery task adaptation from speech to motor imagery and vice-versa is not yet explored, where model pre-trained on one type of mental imagery task is used to fine-tune the target model of other type of mental imagery task. Also, fine-tuning requires all parameters of the pre-trained model to be trained while optimization on target domain data. Nowadays, the size of the models keep getting bigger to attain human brain level intelligence. Therefore, full fine-tuning of such large models becomes compute and space extensive. In recent literature of NLP and CV, parameter efficient fine-tuning methods such as LoRA \cite{hu2021lora}, are proposed that requires only a small number of trainable parameters compared to full fine-tuning while also maintaining the performance. Inspired by a similar work \cite{liu2024dora}, we propose an ensemble of weight-decomposed low rank adaptation method for EEG-based mental imagery task classification.

\section{Methods}

\subsection{Definitions and Notations}
We denote $D_{d_1}^S = \{(X_{d_1}^i, y_{d_1}^i) \mid X_{d_1}^i \in {R}^{N_{c_{1}} \times N_{t_{1}}}$, $y_{d_1}^{i} \in Y_{d_1}\}$ as the source domain, where \(X_{d_1}^i\) is a pre-training EEG dataset trial in the source domain with \(N_{c_{1}}\) spatial channels and \(N_{t_{1}}\) temporal sampling points, and \(Y_{d_1} = \{0: {out}, 1: {in}, 2: {up}\}\) is the label set if \(d_1\) is SI dataset and $Y_{d_1}$ = \{0: \textit{Left Hand}, 1: \textit{Right Hand}, 2: \textit{Both Feet}, 3: \textit{Tonque}\} if \(d_1\) is MI dataset.

Similarly, the target domain is defined as $D_{d_2}^T = \{(X_{d_2}^i, y_{d_2}^i) \mid X_{d_2}^i \in {R}^{N_{c_{2}} \times N_{t_{2}}}$, $y_{d_2}^i \in Y_{d_2}\}$, where \(X_{d_2}^i\) is a fine-tuning EEG dataset trial in the target domain (MI or SI Dataset). 

\subsection{Low-Rank Adaptation}
Low-rank adaptation is the one of the PEFT method that does not change the model architecture and gaining popularity due to its simplicity and efficiency. LoRA (Low-Rank Adaptation) strategy utilizes a simple design that brings practical benefits to dense layers in deep learning models \cite{hu2021lora}. The experimental focus of this paper was primarily on Transformer language models, however, the principles can be applied to other models.

Neural networks often incorporate dense layers that perform matrix multiplications using fully-ranked weight matrices. Inspired by the concept that pre-trained language models function within a low hidden dimension, the authors of this paper proposed that weight updates during model adaptation also exhibit a low intrinsic rank. Consider a pre-trained weight matrix \( W_0 \) in $R^{d \times k}$. The low-rank decomposition of this weight matrix is given by:
\begin{equation}
    W_0 + \Delta W = W_0 + BA
\end{equation}

where  \( B \) $\in$ $R^{d \times r}$ ,  \( A \) $\in$ 
$R^{r \times k}$, and the rank \( r \) satisfies \( r \leq \min(d, k) \). During training, \( W_0 \) is held constant, with only \( A \) and \( B \) (which contain the trainable parameters) being updated. This yields a modified forward pass equation:
\begin{equation}
    h = W_0x + \Delta Wx = W_0x + BAx
\end{equation}

If $r$ is chosen as a very small number, the number of trainable parameter reduces significantly. LoRA approach is resource efficient and have low inference latency. LoRA can be applied to any subset of weight matrices within a neural network to minimize the number of trainable parameters.

\begin{figure}[t!] 
   \centering
   \includegraphics[width=1\textwidth]{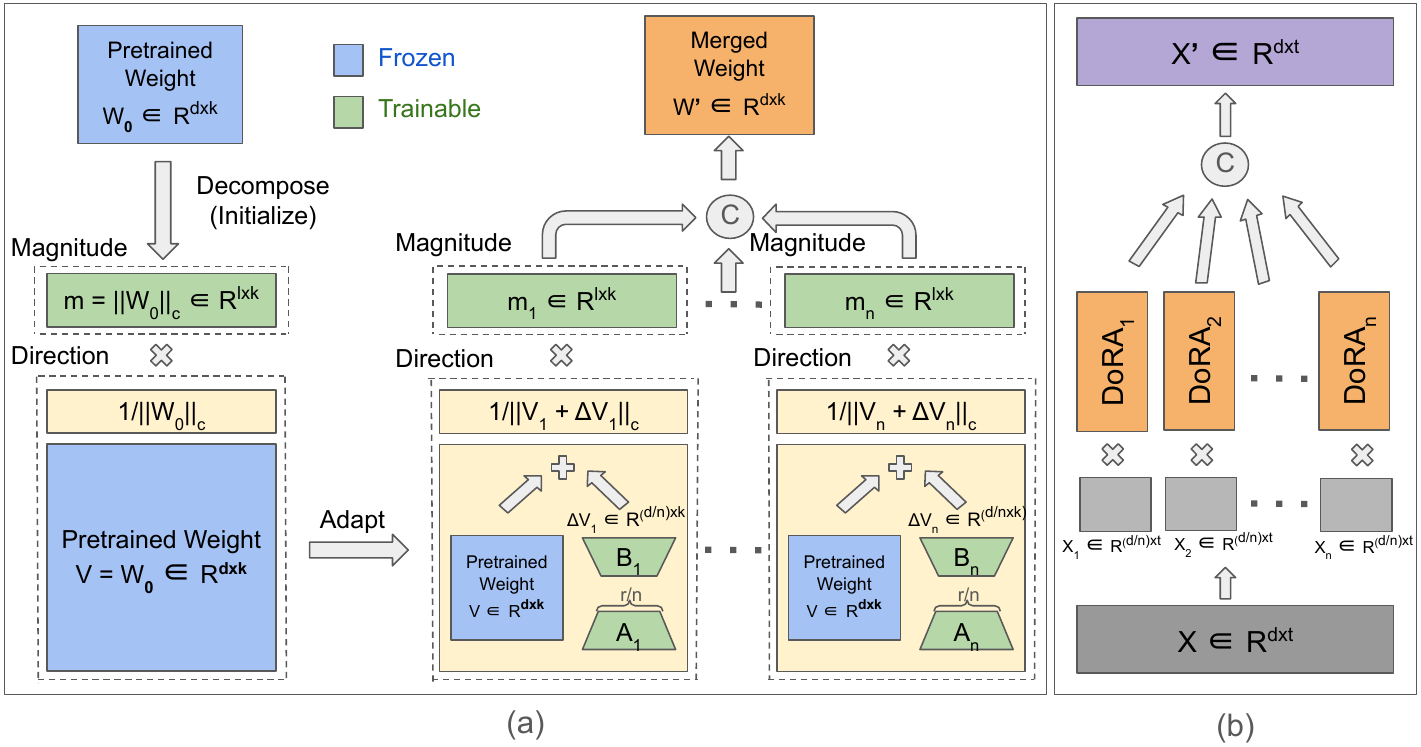} 
   \caption{(a) Overview of proposed EDoRA paramter-efficient fine-tuning approach. It depicts the overall parameter updation process of EDoRA, before and after fine-tuning (b) Feature updation via EDoRA adaptation. [$\copyright$ symbol represents concatenation, $\times$ symbol represents product, $X$ represents input features, $X^{'}$ represents output features.] } 
   \label{fig:overview} 
\end{figure}

\subsection{Proposed method}

The proposed method EDoRA is inspired by the weight-decomposed low-rank adaptation method (DoRA) \cite{liu2024dora}. This method enhances efficiency by dissecting pre-trained model weights into magnitude and directional constituents, each of which is subsequently fine-tuned for optimized performance. This decomposition allows for concentrated updates that minimize the quantity of trainable parameters while maximizing the efficacy of the learning process. The adaptation process is defined in the following steps:

\textit{Initial Decomposition:} The process starts with the division of pre-trained model weights into magnitude and directional components. This separation is crucial for enabling specific updates during fine-tuning.
The weight decomposition is represented as:
\begin{equation}
      W = m \frac{V}{\|V\|_c}  =  \|W\|_c \frac{W}{\|W\|_c}
\end{equation}

where \( W \) is the weight matrix, \( m \) is a vector representing the magnitude, and \( V \) is the matrix representing directional values, with \( \|V\|_c \) denoting the column-wise norm \cite{liu2024dora}. The number of parameters in the directional component is more than the magnitude component; therefore, only the directional component is fine-tuned via LoRA to keep the number of trainable parameters efficient \cite{liu2024dora}.

\textit{Fine-tuning of Directional Component:} The directional component is fine-tuned using Low-Rank Adaptation (LoRA) \cite{hu2021lora}. This process involves updates focused primarily on parameters that offer the highest utility, thus enhancing efficiency.
The adapted weight \( W' \) after fine-tuning is given by:
\begin{equation}
      W' = m \frac{V + \Delta V}{\|V + \Delta V\|_c}  =  m \frac{W_0 + BA}{\|W_0 + BA\|_c} 
\label{eq:equation_4}
\end{equation}

where $V$ represents the weights of pre-trained model, and  \( \Delta V \) is the change in directional component learned by multiplying two low-rank matrices \( B \) and \( A \) \cite{liu2024dora}.

\begin{figure}[t!] 
   \centering
   \includegraphics[width=1\textwidth]{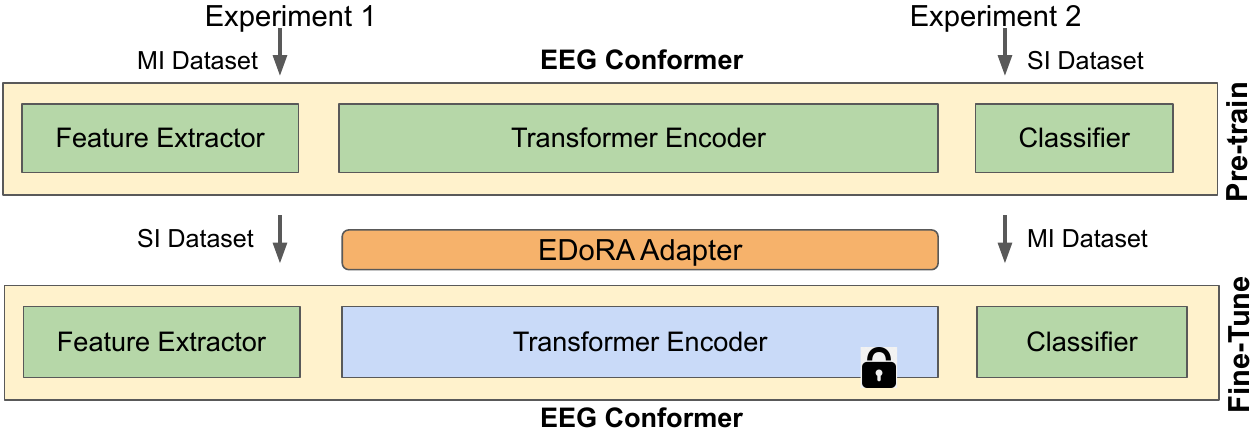} 
   \caption{Framework of the proposed method. Two experiments are performed in this work, and in these experiments EEG Conformer model is pre-trained on one dataset, and then fine-tuned on other dataset with only EDoRA adapter on each operation of transformer encoder of EEG Conformer and vice-versa. [Freezed weights are shown with lock]  } 
   \label{fig:experiment_procedure} 
\end{figure}

Eq. \ref{eq:equation_4} represents the fine-tuned weights of the DoRA method. As EEG signals are highly non-stationary, the features learned through fine-tuning individual DoRA adapters have a high variance in feature learning that may result in overfitting and local optima. Therefore, an ensemble of multiple DoRA adapters reduces the overfitting and local optima issues, which leads to stable and reliable performance \cite{chakladar2021eeg}. Therefore, the proposed method EDoRA is an ensemble of multiple DoRA adapters. The weight updation in each individual adapter of EDoRA will follow the Eq. \ref{eq:equation_4}. The input feature $x \in R^{d \times t}$ is split into $n$ equal parts that gives $x_i \in R^{d \times \frac{t}{n}}$, where $i=0,1,2 \dots n$. Each $x_i$ is fed to $DoRA_i$, and the output weights of each $DoRA_i$ are concatenated. This process is depicted in Fig. \ref{fig:overview} (b). The modified forward pass equation for EDoRA is defined by the following equations:

\begin{equation}
\begin{array}{l} 
      h = concat^{n}_{i=0} \Big (m_i \frac{V x_i + \Delta V_i x_i}{\|V x_i + \Delta V_i x_i \|_c} \Big ) \\
      \hspace{3mm} = concat^{n}_{i=0} \Big (m_i \frac{W_0 x_i + (B_i A_i) x_i}{\|W_0 x_i + (B_i A_i) x_i \|_c} \Big )
\end{array}       
\label{eq:equation_5}
\end{equation}

where $n$ denotes number of equal parts of the input features $x$, \(A_i\) $\in$  ${R}^{\frac{r}{n} \times d}$ and \(B_i\) $\in  {R}^{k \times \frac{r}{n}}$ are the low rank matrices with rank $\frac{r}{n}$ , $x_i \in R^{d \times \frac{t}{n}}$ is the input feature matrix, $m_i \in 1 \times k$ is the magnitude vector, $W_0 \in R^{d \times k}$ are the pre-trained model weights, and $h \in R^{k \times t} $ represents the updated feature.

Detailed overview of weight updation in the proposed method is depicted in Fig \ref{fig:overview} (a), and detailed depiction of feature updation is given in Fig. \ref{fig:overview} (b).

\subsection{Optimization Procedure}

In this study, the EEG Conformer model is firstly pre-trained on \(d_1\) EEG data in \(D_{d_1}^S\) and then fine-tuned using \(d_2\) EEG data in \(D_{d_2}^T\). The decoding model can be represented as a classifier \(m: {R}^{N_c \times N_t} \to y_{d_1} \mid y_{d_2}\) which is defined as:

\begin{equation}
    m(X_i; \theta) = g(\phi(X_i; \theta_\phi);\phi_g)
\end{equation}

where \(\phi\) denotes feature extraction and transformer encoder module with parameters \(\theta_\phi\), and $g$ denotes classifier module with learnable parameters $\phi_g$. This model learns the classification of data by minimizing the prominently used cross-entropy loss \cite{lotey2022cross}. Fig. \ref{fig:experiment_procedure} illustrates the network architecture and adaptation strategy of our work.

\section{Experimental Results}

\subsection{Datasets}

\subsubsection{Motor Imagery Dataset}

The BCI Competition IV 2a is a publicly available motor imagery dataset that comprises EEG data from 9 subjects performing motor imagery tasks of the left hand, right hand, both feet, and tongue \cite{tangermann2012review}. The dataset features EEG signals recorded at a 250 Hz sampling rate from 22 electrodes over two sessions, each containing six runs of 48 trials per motor task, totaling 288 trials per session. 
Hereon, this dataset will be depicted as '\textit{MI dataset}'.
\subsubsection{Speech Imagery Dataset}
The publicly available Arizona state university (ASU) dataset is used for the classification of the speech imagery tasks \cite{nguyen2017inferring}. In this study, we used a dataset for short word classification. The dataset consists of three class SI of the English words “in”, "out" and “up.” Each class consists of 100 trials, and a single trial lasts for 5 s. The data were acquired from six subjects with 60 EEG channels. The data was pre-processed using a frequency range of 8–70 Hz.
Hereon, this dataset will be depicted as '\textit{SI dataset}'.
\subsubsection{Data pre-processing}
MI dataset is kept at 250Hz sampling rate whereas SI dataset is downsampled from 1000Hz to 250Hz. Keeping the data temporal length same, $4$ second data is being used for both datasets. Further, $4$th-order Butterworth filter is used to eliminate low-frequency noise in the $4$–$40$ Hz range and z-score normalization applied.
 
\subsection{Experimental Details}

To implement the proposed and compared models, PyTorch deep learning library \cite{paszke2019pytorch} was used and executed on a NVIDIA RTX Quadro 5000 GPU system with $16$ GB GPU memory and $16$ GB of RAM. For pre-training, the model is trained for $2000$ epochs with learning rate of $0.0002$ and $80$-$20$ split is used for both datasets, i.e., $80\%$ training data and $20\%$ testing data. For fine-tuning, all methods are trained for $500$ epoch having batch size of $72$. In this work, we have followed the settings of the dataset split most often used in the literature. $80$-$20$ split is used for SI dataset whereas session based split used in the original paper is employed for MI dataset. For both pre-training and fine-tuning, Adam optimizer is used with constant learning rate of $0.0002$ with $\beta_1$ as $0.5$ and $\beta_2$ as $0.999$. The metric used to evaluate the model performance is accuracy, confusion matrix, AUC-ROC score and kappa measure \cite{kaushik2023motor,lotey2022cross}.

\begin{table}[!t]
\centering
\caption{Accuracy comparison of proposed method with full fine-tuning and other parameter efficient adaptation methods for SI and MI EEG signal classification. [Standard deviation is reported in round brackets.]}
\begin{tabular}{c|c|cccc}
\hline
\multirow{2}{*}{\textbf{Dataset}}  &   \multirow{2}{*}{\textbf{Subject}}      & \multicolumn{4}{c}{\textbf{Methods}}          \vspace{1pt}  \\ \cline{3-6}
\multicolumn{1}{l|}{} &  & \textbf{Fine-Tune} & \textbf{LoRA} \cite{hu2021lora} & \textbf{DoRA} \cite{liu2024dora} & \textbf{EDoRA(our)} \vspace{1pt} \\ \hline 
\multirow{7}{*}{Speech Imagery}  & 1       & 51.67 & 50.00 & 50.00 & 53.33          \\ 
                      & 3       & 48.33 & 50.00 & 48.33 & 41.67          \\ 
                      & 5       & 48.33 & 45.00 & 45.00 & 50.00          \\ 
                      & 6       & 51.67 & 48.33 & 50.00 & 53.33          \\ 
                      & 8       & 58.33 & 65.00 & 65.00 & 61.67         \\
                      & 12      & 51.67 & 53.33 & 53.33 & 58.33          \\ \cline{2-6}
                      & Average & 51.67 (3.65) & 51.94 (6.95) & 51.94 (6.95) & \textbf{53.06} (6.94) \\ \hline
\multirow{10}{*}{Motor Imagery} & 1       & 81.60 & 79.86 & 79.51 & 81.94          \\ 
                      & 2       & 52.78 & 51.74 & 52.08 & 49.31         \\ 
                      & 3       & 85.76 & 84.38 & 84.38 & 86.11          \\ 
                      & 4       & 65.63 & 65.63 & 66.67 & 67.71         \\ 
                      & 5       & 42.36 & 44.44 & 44.44 & 44.79          \\ 
                      & 6       & 51.04 & 51.04 & 52.43 & 56.25          \\ 
                      & 7       & 81.25 & 78.13 & 78.13 & 80.90          \\ 
                      & 8       & 80.21 & 78.47 & 78.82 & 80.21          \\ 
                      & 9       & 79.86 & 72.92 & 73.96 & 78.13          \\ \cline{2-6}
                      & Average & 68.94 (16.36) & 67.40 (14.81) & 67.82 (14.61) & \textbf{69.48} (15.60) \\ \hline
\end{tabular}
\label{table:comparison}
\end{table}

\subsection{Performance Evaluation}
The performance of the proposed parameter-efficient adaptation methods is validated through two mental imagery task datasets, i.e., the speech imagery dataset (SI dataset) and the motor imagery dataset (MI dataset). The method is first pre-trained on one dataset and then adapted to the other dataset via fine-tuning and vice-versa. The evaluation performance of the each fine-tuned dataset is discussed in the following sections of this paper. 

The performance of the proposed parameter-efficient adaptation method is stated in Table \ref{table:comparison}. The first method compared with the proposed method is a fully fine-tuned model (depicted as fine-tuned in the table). This method requires training all parameters of the pre-trained model to optimize the target domain data. The other compared methods are state-of-the-art (SOTA) parameter-efficient adaptation methods, LoRA \cite{hu2021lora} and DoRA \cite{liu2024dora}. The comparison table of accuracy metrics validates the effectiveness of the proposed adaptation method as our method has performed better than full fine-tuning, which requires a huge number of parameters to train, whereas the proposed method only requires a small number of the total trainable parameters. 

\begin{figure}[!t]
        \centering

        \begin{subfigure}[t]{0.47\textwidth}  
            \centering 
            \includegraphics[scale=0.25]{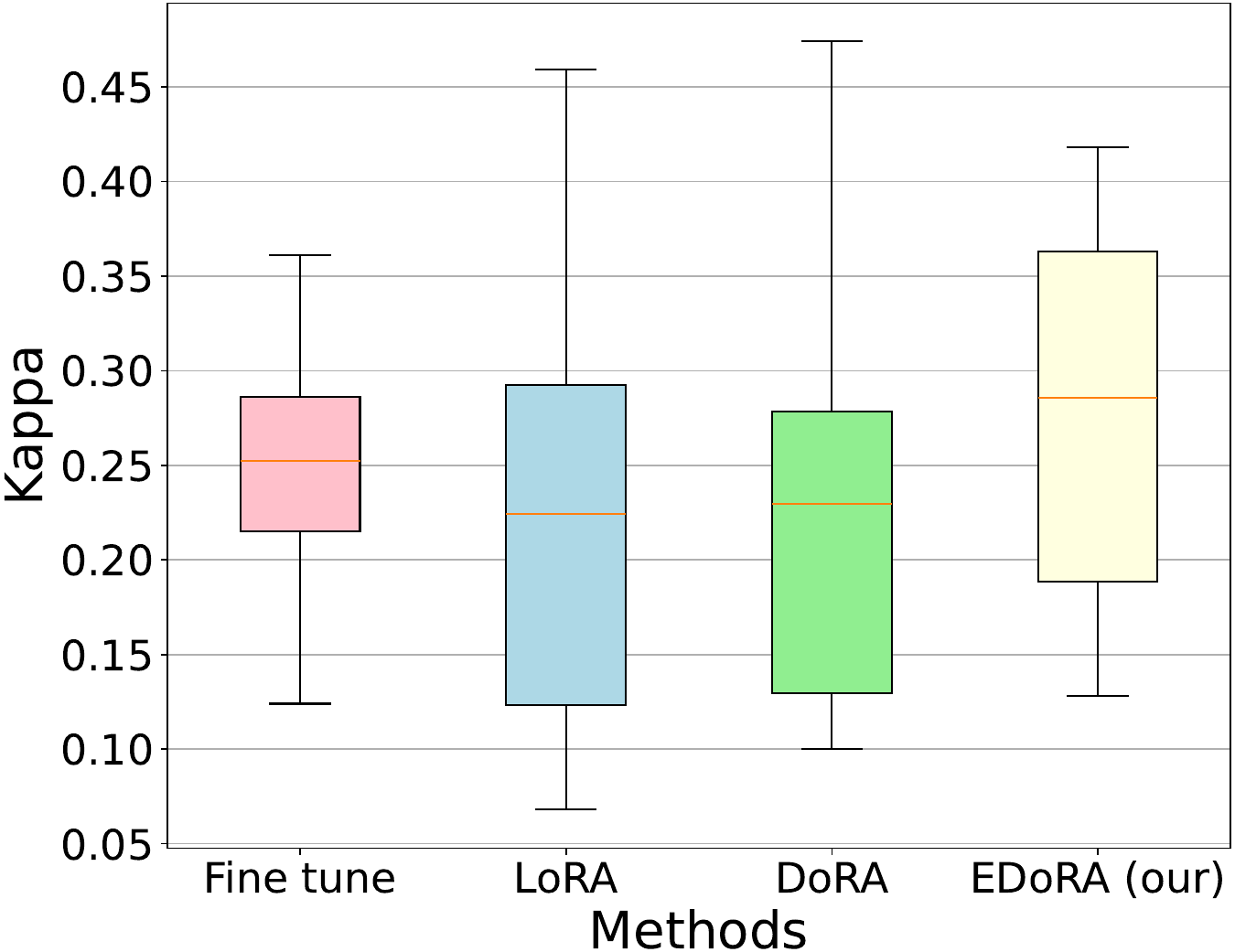}
            \caption[]%
            {SI Dataset}    
            \label{fig:cm_aus_vowel}
        \end{subfigure}%
        ~
        \begin{subfigure}[t]{0.47\textwidth}  
            \centering 
            \includegraphics[scale=0.25]{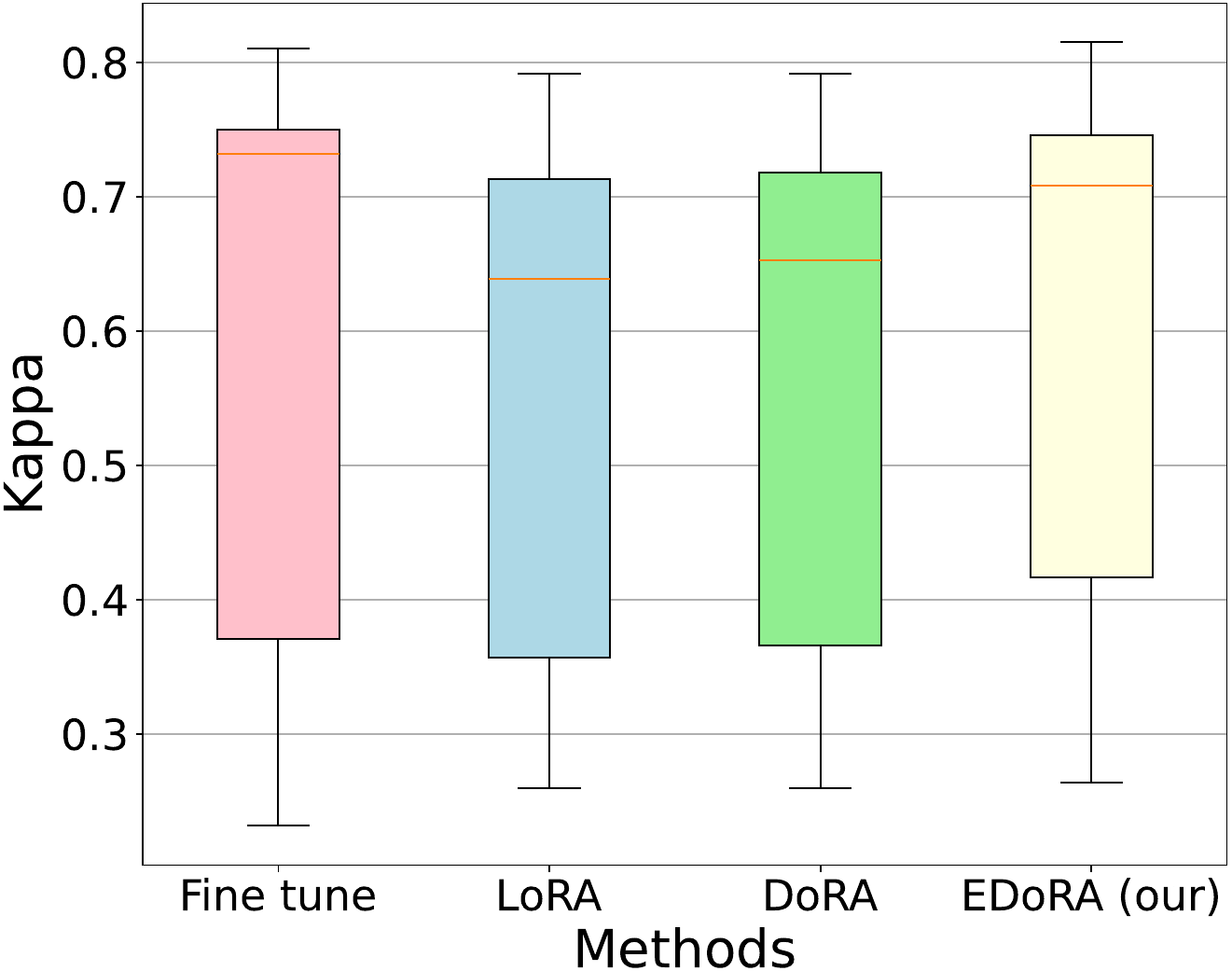}
            \caption[]%
            {MI Dataset}    
            \label{fig:cm_aus_short}
        \end{subfigure}%

        \caption[  ]{Mean Kappa measure of proposed method and compared methods.} 
        \label{fig:kappa_measure}
\end{figure}

In the SI dataset, the accuracy of the proposed method is $1.39\%$ more than full fine-tuning and $1.12\%$ more than both LoRA and DoRA. Similarly, in the MI dataset, the proposed method is superior to the compared methods with margins of $0.54\%$, $2.08\%$, and $1.66\%$ than full fine-tuning, LoRA and DoRA, respectively. Fig. \ref{fig:kappa_measure} shows the kappa value comparison of the proposed method with compared methods. The box plots show that for the SI dataset, the median is clearly distant from the compared methods, and the minimum value of the kappa measure is also better than all of the compared methods. For the MI dataset, although the median of the proposed method is below the median of the full fine-tune method, the box length is smaller, and the minimum value of the kappa is higher. Thus, the kappa measure of both datasets suggests that the performance of the proposed method is better than the compared methods in both datasets.

\begin{figure}[!t]
        \centering

        \begin{subfigure}[t]{0.48\textwidth}  
            \centering 
            \includegraphics[scale=0.07]{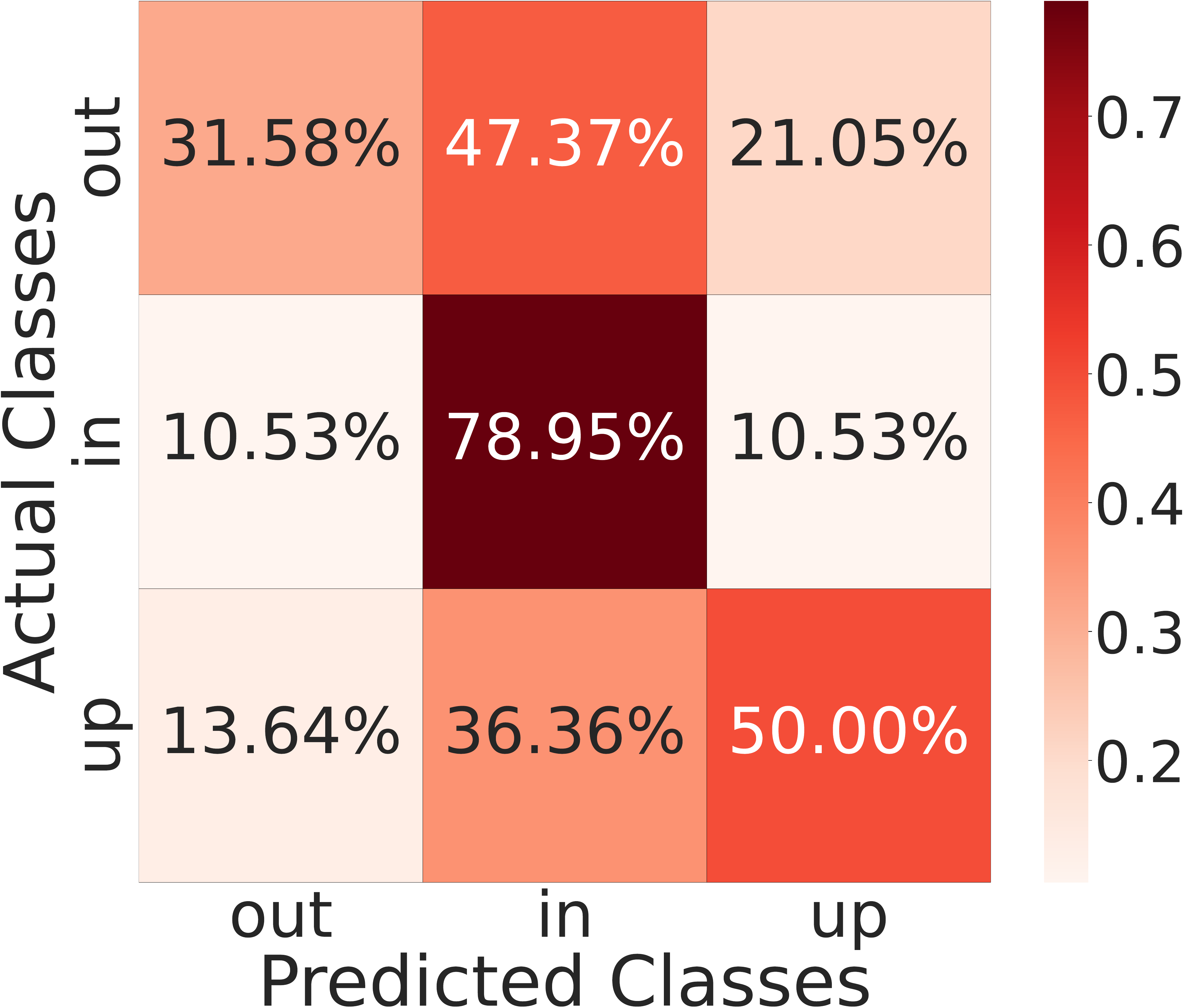}
            \caption[]%
            {SI Dataset Subject 6}    
            \label{fig:cm_si_6}
        \end{subfigure}%
        \begin{subfigure}[t]{0.48\textwidth}  
            \centering 
            \includegraphics[scale=0.07]{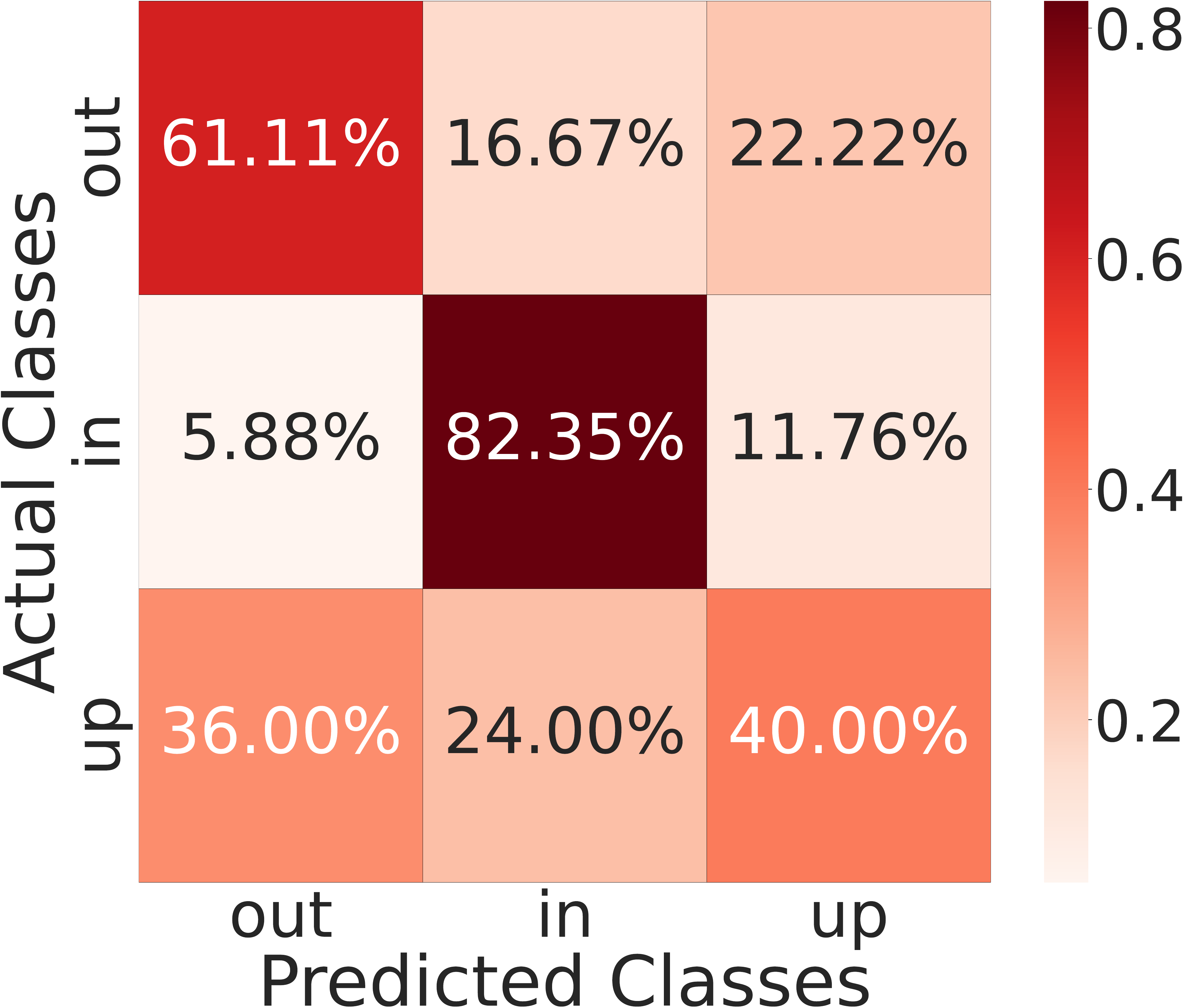}
            \caption[]%
            { SI Dataset Subject 12}    
            \label{fig:cm_si_12}
        \end{subfigure}%
        \\
        \begin{subfigure}[t]{0.48\textwidth}  
            \centering 
            \includegraphics[scale=0.08]{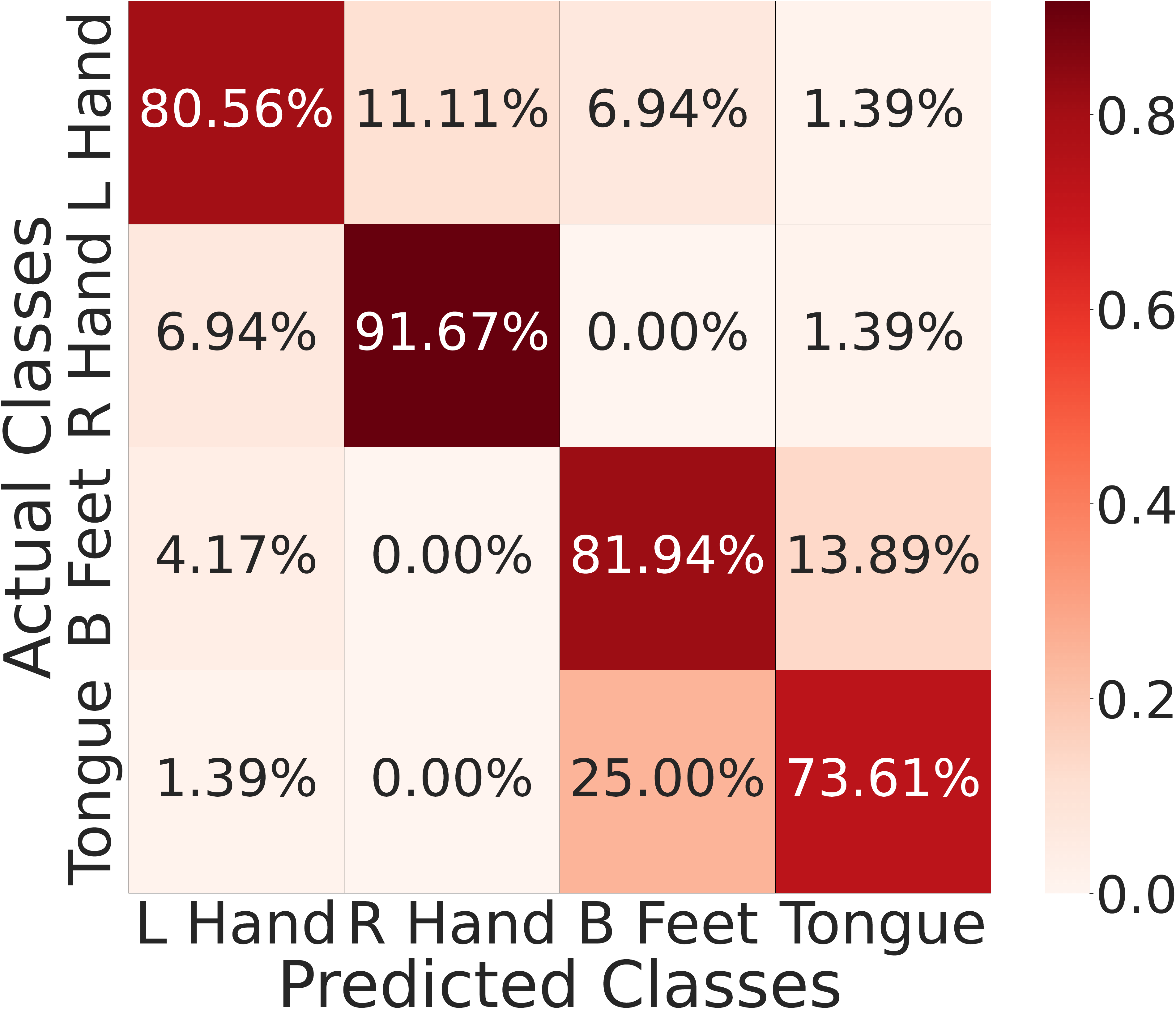}
            \caption[]%
            {MI Dataset Subject 1}    
            \label{fig:cm_mi_1}
        \end{subfigure}%
         ~ 
        \begin{subfigure}[t]{0.48\textwidth}  
            \centering 
            \includegraphics[scale=0.08]{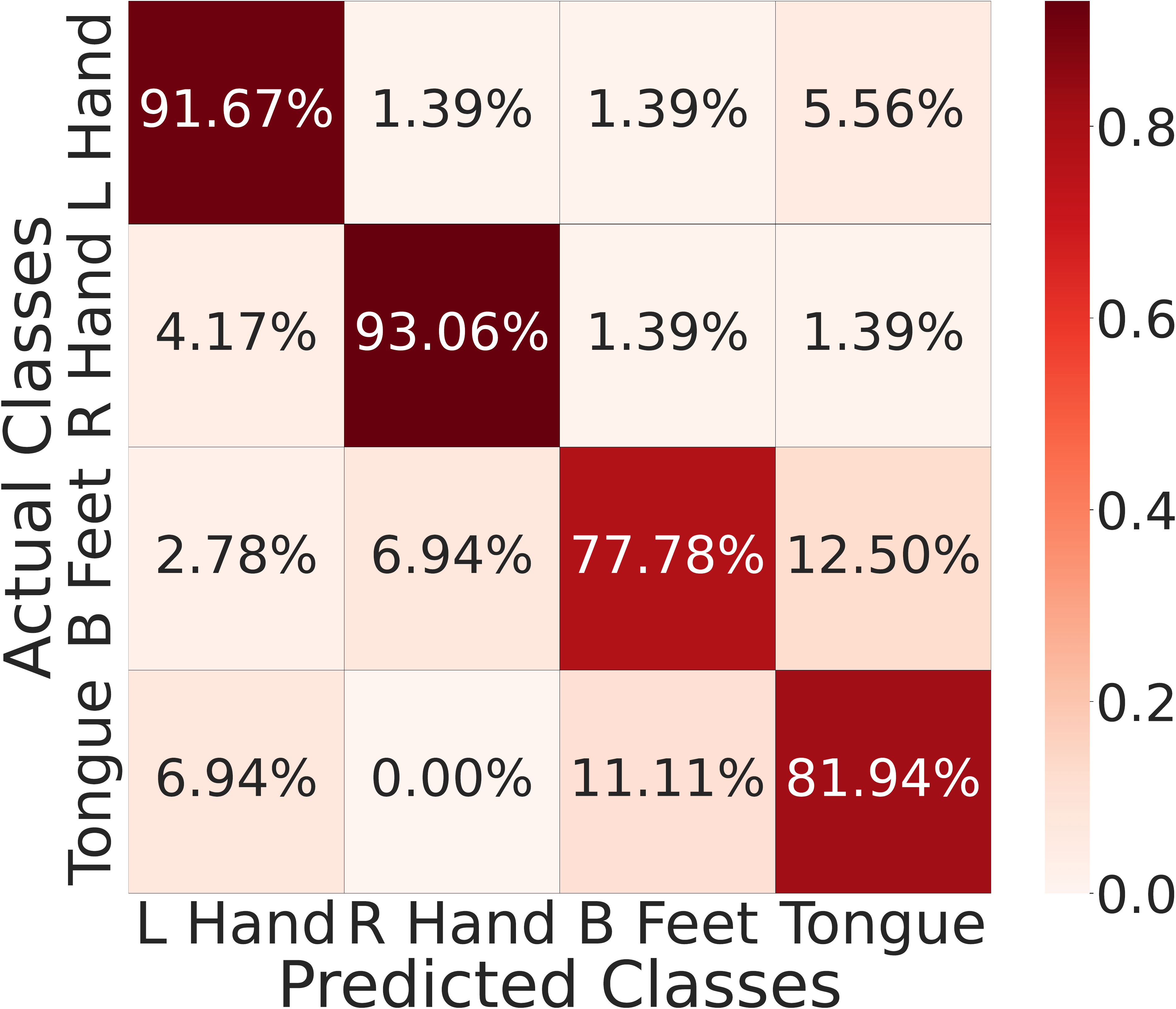}
            \caption[]%
            {MI Dataset Subject 3}    
            \label{fig:cm_mi_3}
        \end{subfigure}%

        \caption[  ]{Confusion matrices of proposed method on two subjects of SI and MI dataset.} 
        \label{fig:confusion_matrices}
\end{figure}

\begin{figure}[!th]
        \centering
        \begin{subfigure}[t]{0.49\textwidth}  
            \centering 
            \includegraphics[scale=0.4]{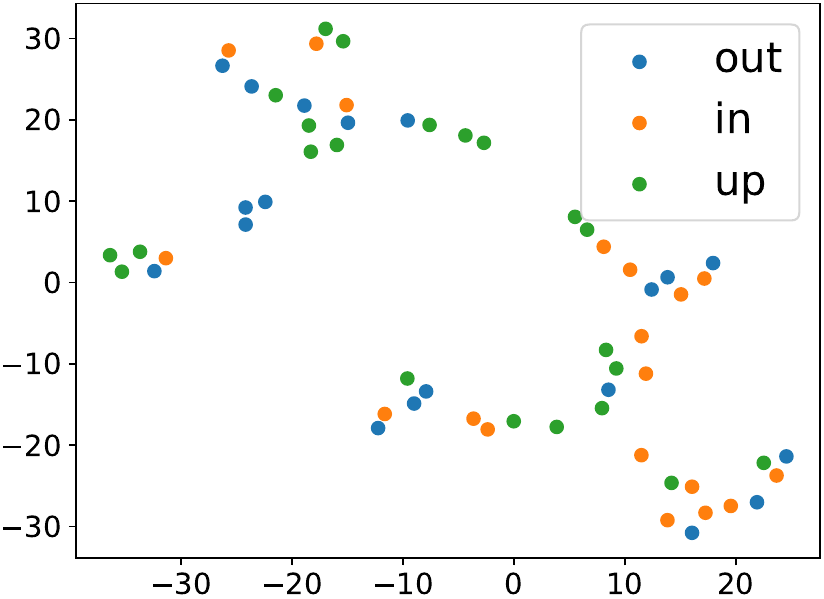}
            \caption[]%
            {SI Dataset Subject 6 }    
            \label{fig:tnse_si_6}
        \end{subfigure}%
        ~  
        \begin{subfigure}[t]{0.49\textwidth}  
            \centering 
            \includegraphics[scale=0.4]{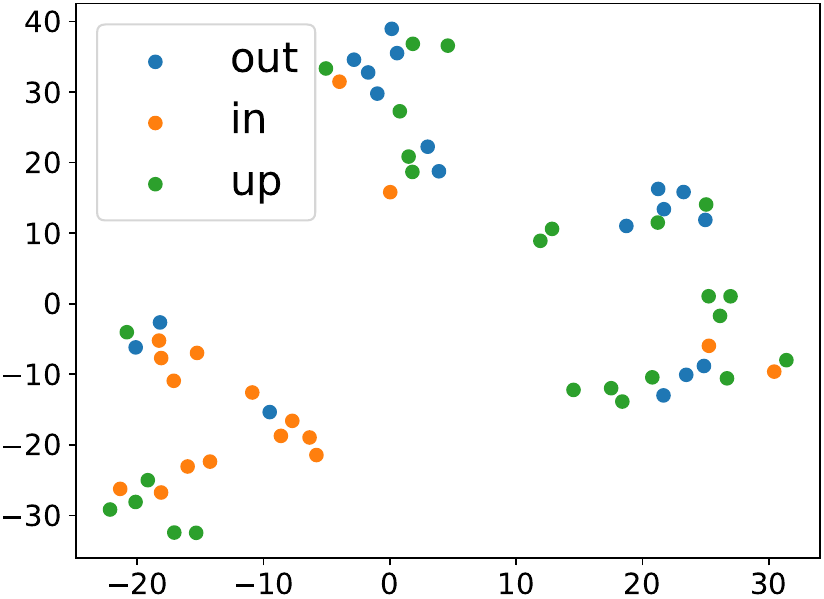}
            \caption[]%
            { SI Dataset Subject 12}    
            \label{fig:tsne_si_12}
        \end{subfigure}%
        \\
        \begin{subfigure}[t]{0.49\textwidth}  
            \centering 
            \includegraphics[scale=0.4]{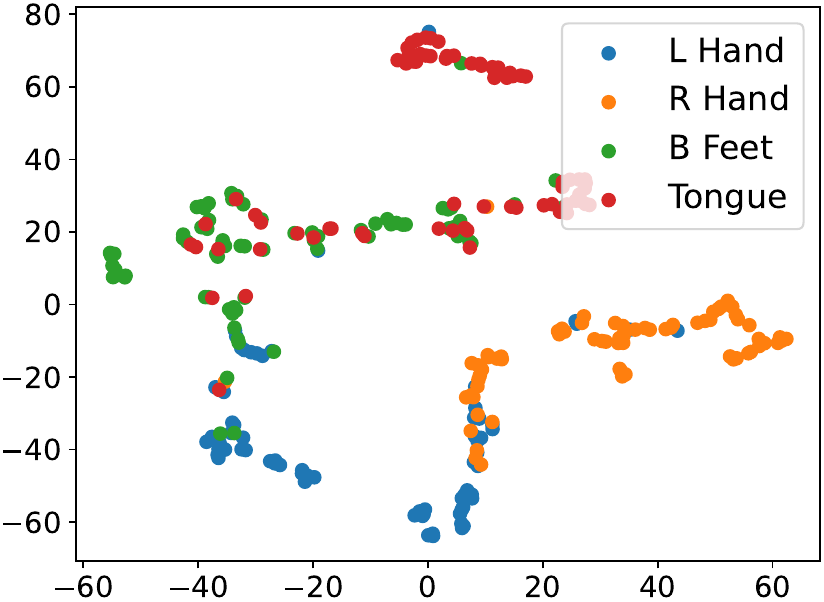}
            \caption[]%
            {MI Dataset Subject 1 }    
            \label{fig:tsne_mi_1}
        \end{subfigure}%
        ~
        \begin{subfigure}[t]{0.49\textwidth}  
            \centering 
            \includegraphics[scale=0.4]{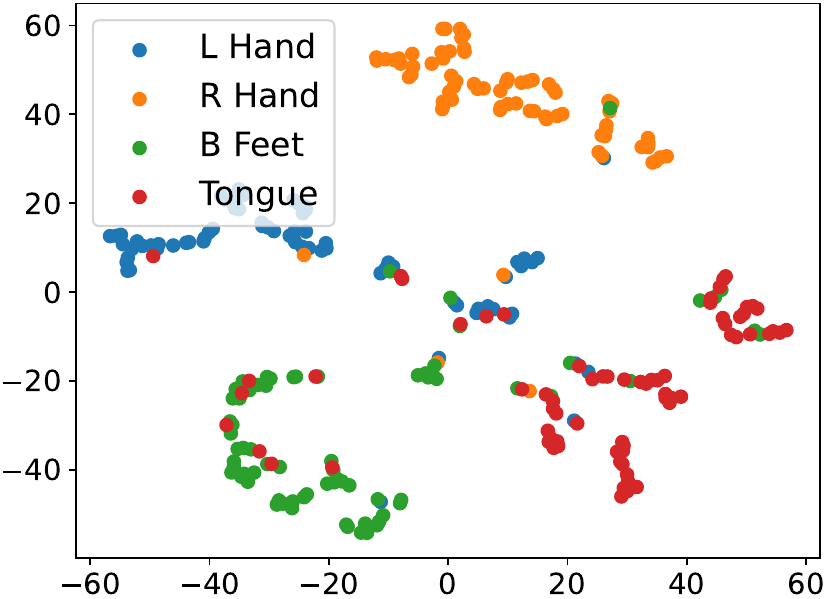}
            \caption[]%
            {MI Dataset Subject 3 }    
            \label{fig:tsne_mi_3}
        \end{subfigure}%
        \\
        \caption[  ]{t-SNE plots of proposed method (EDoRA) on two subjects of SI and MI dataset.} 
        \label{fig:tsne_plots}
\end{figure}

Fig. \ref{fig:confusion_matrices} and Fig. \ref{fig:tsne_plots} show the per-class discriminative power of the proposed method. Fig. \ref{fig:confusion_matrices} shows the confusion matrices of two subjects of each dataset. Fig. \ref{fig:cm_si_6} and Fig. \ref{fig:cm_si_12} show that class "in" is more accurately classified compared to the other two classes. The reason might be the similarity of the "out" and "up" words and the dissimilarity of "in" from these two words. Also, the behavior is consistent on the confusion matrices of both subjects. Fig. \ref{fig:cm_mi_1} and Fig. \ref{fig:cm_mi_3} show that for motor imagery tasks, the per-class classification performance is balanced. However, left-hand imagined movement and right-hand imagined movement are least confused with other classes by the proposed methods. The class-wise clustering of the EEG signals in the feature space is shown with the help of t-SNE plots in Fig. \ref{fig:tsne_plots}. Fig. \ref{fig:tnse_si_6} and Fig. \ref{fig:tsne_si_12} show where the testing data points of the SI dataset lie in the feature space. These figures show the clusters of three classes of the SI dataset, where the cluster of the "in" class is clear and distinct, whereas the "out" and "up" classes show some overlap. This finding is similar to the confusion matrices of the SI dataset. Fig. \ref{fig:tsne_mi_1} and Fig. \ref{fig:tsne_mi_3} depict the t-SNE plots for two subjects of the MI dataset, and these plots show clear and distinct clusters of all four classes of the MI dataset. Therefore, confusion matrices and t-SNE plots show the classification performance of the proposed parameter-efficient adaptation method.

\begin{table}[!t]
\centering
\caption{Accuracy comparison of proposed method with full fine-tuning and SOTA parameter-efficient adaptation methods with different ranks (r) for SI and MI datasets. [D: Dataset, Standard deviation is reported in round brackets, EDora\textsuperscript{$\dagger$} is proposed method with n=4.]}
\begin{tabular}{c|c|c|c|c|c|c}
\hline
\multirow{2}{*}{\textbf{D}} & \multicolumn{1}{l|}{\multirow{2}{*}{\textbf{Rank}}} & \multicolumn{5}{c}{\textbf{Methods}}                                                                                                            \\ \cline{3-7}
                         & \multicolumn{1}{l|}{}                      & \textbf{Fine-tune}                                         & \textbf{LoRA}          & \textbf{DoRA}          & \textbf{EDoRA(our)}    & \textbf{EDora\textsuperscript{$\dagger$}(our)} \\ \hline
\multirow{4}{*}{SI}     & 2                                         & \multicolumn{1}{c|}{\multirow{4}{*}{51.67 (3.65)}} & 52.22 (6.47)  & 52.22 (6.64)  & \textbf{53.33} (8.63)  & $-$             \\
                         & 4                                         & \multicolumn{1}{c|}{}                              & 51.94 (6.95)  & 51.94 (6.95)  & \textbf{53.06} (6.94)  & 51.95 (6.87)                       \\
                         & 8                                         & \multicolumn{1}{c|}{}                              & 50.56 (6.12)  & \textbf{51.39} (6.78)  & \textbf{51.39} (5.10)  & 50.83 (7.21)                       \\
                         & 16                                        & \multicolumn{1}{c|}{}                              & 52.50 (7.28)  & 51.11 (6.38)  & 51.94 (5.42)  & \textbf{52.78} (4.55)                       \\ \hline
\multirow{4}{*}{MI}     & 2                                         & \multirow{4}{*}{68.94 (16.36)}                    & \textbf{68.90} (15.81) & 68.75 (15.58) & 68.63 (15.27) & $-$              \\
                         & 4                                         &                                                   & 67.40 (14.81) & 67.82 (14.61) & \textbf{69.48} (15.60) & 68.09 (14.34)                      \\
                         & 8                                         &                                                   & 68.36 (15.80)  & 67.94 (16.20) & \textbf{68.94} (14.50) & 69.52 (13.08)                      \\
                         & 16                                        &                                                   & 66.63 (14.79) & 67.48 (14.66) & 67.94 (14.30) & \textbf{68.83} (14.82)  \\  \hline                 
\end{tabular}
\label{table:rank_comparison}
\end{table}

The parameter-efficient adaptation methods have a hyper-parameter of rank that decides the number of neurons of the adapters. Hence, we present the comparison of the performance of the proposed parameter-efficient method EDoRA with different rank values. Table \ref{table:rank_comparison} shows the accuracy metric of the proposed EDoRA method and compared methods. In this table, for each rank, the highest accuracy measure among LoRA, DoRA, and EDoRA (our) is shown in bold. For both datasets, EDoRA has performed superior to compared parameter-efficient adaptation methods with the exception of rank 2, where LoRA ($68.90\%$) performed slightly better than EDoRA ($68.63\%$). However, maximum accuracy with the least standard deviation is attained by EDoRA only (rank=8, n=4, accuracy=$69.52\%$). Additionally, the highest accuracy of the proposed method is better than the accuracy of the full fine-tune method. It shows the strength of the low-rank adaptation approach with a very small number of parameters over full fine-tuning. These results demonstrate the ability of EDoRA to adapt from one mental imagery dataset to another mental imagery dataset and vice versa.

\subsection{Ablation study}

\begin{figure}[!t]
        \centering
        \begin{subfigure}[t]{0.5\textwidth}
            \hspace{-9mm}
            \centering 
            \includegraphics[width=\linewidth]{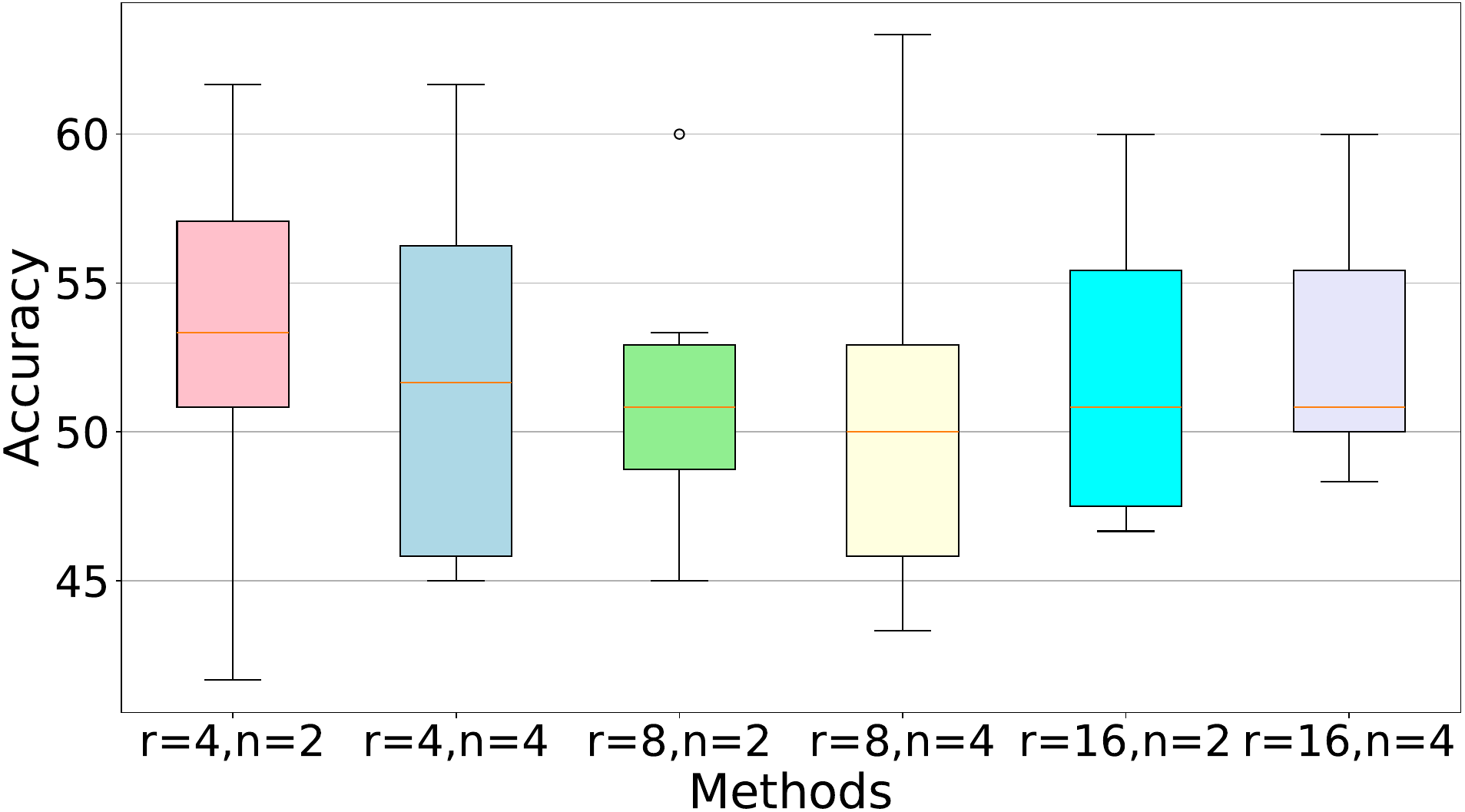}
            \caption[]%
            {SI Dataset}    
            \label{fig:ablation_si_r_n}
        \end{subfigure}%
        ~
        \begin{subfigure}[t]{0.5\textwidth}  
            \hspace{-9mm}
            \centering 
            \includegraphics[width=\linewidth]{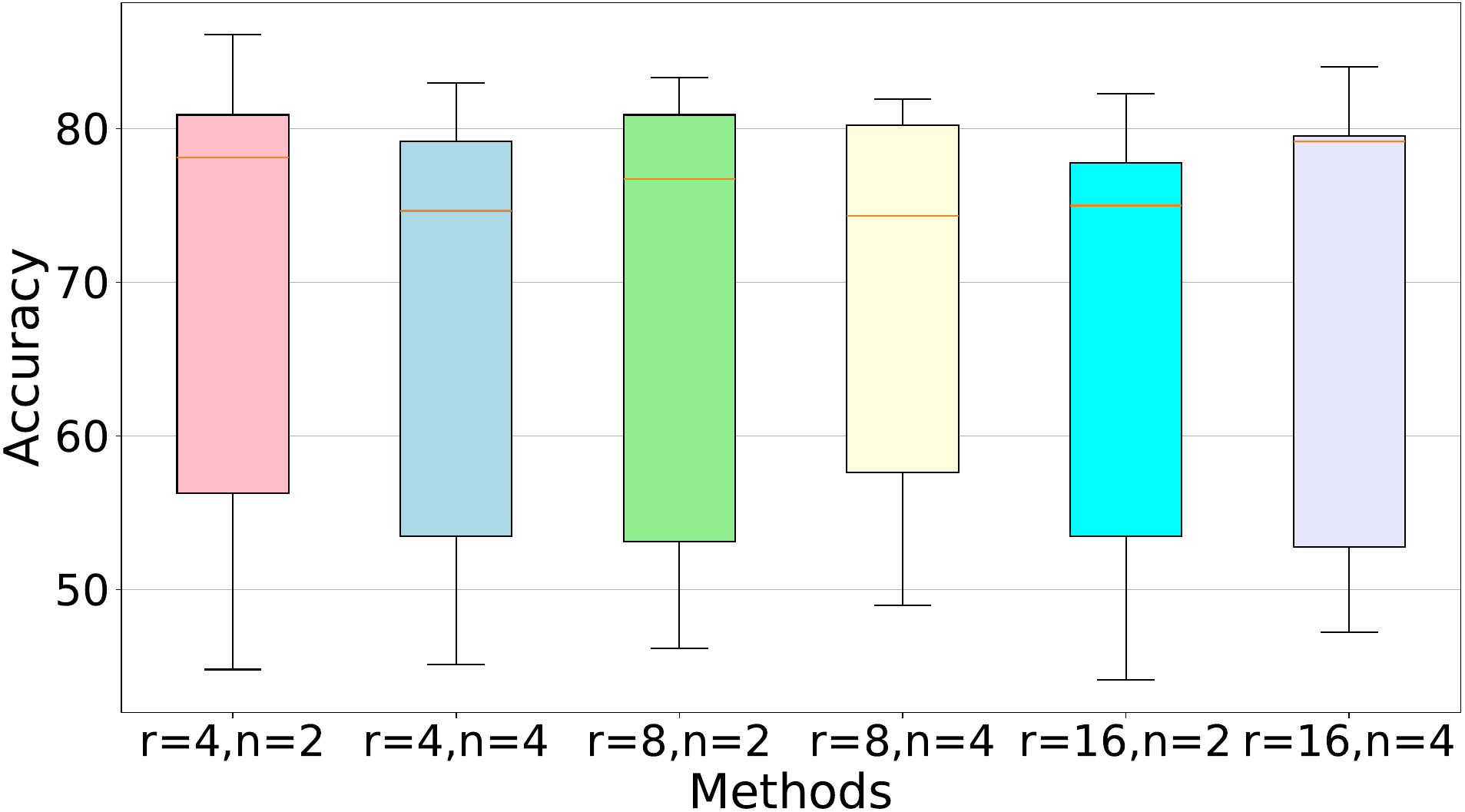}
            \caption[]%
            {MI Dataset}    
            \label{fig:ablation_mi_r_n}
        \end{subfigure}%

        \caption[  ]{Accuracy of proposed method (EDoRA) with different ranks (r) and segment (n).} 
        \label{fig:ablation_r_n}
\end{figure}

In our proposed ensemble of the PEFT adapter method, the number of adapters (n) is a hyper-parameter of the EDoRA method. Therefore, we demonstrate the impact of the number of adapters in an ensemble with different ranks (r) on EDoRA in Fig. \ref{fig:ablation_r_n}. This ablation study shows that for the SI dataset, the performance of all variants is quite variable. This phenomenon can be because the number of samples in the testing set is only $60$. It is a small number, and misclassifying only a few samples can result in lower accuracy. Hence, variants of the SI dataset show a variable performance. However, for the MI dataset, the performance is similar for all of the variants of EDoRA as the number of testing samples is $288$, which is comparatively larger than the testing samples of the SI dataset. Therefore, this analysis suggests the robustness of the proposed method EDoRA over different numbers of n and rank r in both datasets.

\begin{table}[!t]
\centering
\caption{Trainable parameter requirement and mean accuracy of proposed method and compared state-of-the-art parameter-efficient adaptation methods.}
\begin{tabular}{c|c|ccc}
\hline
\textbf{Dataset}     &  \textbf{Method}        & \textbf{LoRA \cite{hu2021lora}} & \textbf{DoRA \cite{liu2024dora}} & \textbf{EDoRA(our)}  \\ \hline
\multirow{2}{*}{Speech Imagery} & \#Parameters &  17k               &  19k          &       21k              \\ 
& Average Accuracy &  51.94 (6.95) &  51.94 (6.95)  & \textbf{53.06} (6.94)         \\ \hline
\multirow{2}{*}{Motor Imagery} & \#Parameters &   17k         &     19k       &            21k        \\ 
& Average Accuracy & 67.40 (14.81)   & 67.82 (14.61) & \textbf{69.48} (15.6)     \\ \hline
\end{tabular}
\label{table:parameters}
\end{table}

\subsection{Parameter analysis}

Table \ref{table:parameters} shows the number of trainable parameters required by EDoRA and compared PEFT methods (LoRA and DoRA). According to this table, the difference in parameters between EDoRA and DoRA is only $2k$, but the difference in accuracy is $1.12\%$ and $1.66\%$ in SI and MI datasets, respectively. A similar behavior is observed when comparing EDoRA with LoRA. It shows that our method is comparable to the other SOTA methods in the trade-off of parameters and performance.

\section{Conclusion}
In this work, we explored the parameter-efficient fine-tuning methods for EEG-based mental imagery task adaptation. Our work is the first to explore the performance and efficiency of parameter-efficient adaptation methods that do not require all parameters of the pre-trained model to be trained. Instead, it trains only a small amount of parameters based on the rank-decomposition technique. We proposed EDoRA, a parameter-efficient fine-tuning method that is an ensemble of multiple parameter-efficient adapters. These adapters decompose the pre-trained weights into magnitude and direction components and adapt these components to the target domain for enhanced fine-tuning. This work is the first to investigate transfer learning-based domain adaptation of speech imagery task from motor imagery task and vice-versa. The effectiveness of the proposed method is validated on two publicly available mental imagery datasets, one dataset of speech imagery and the other of motor imagery. For elevated feature extraction, we have adopted a convolutional transformer approach-based model known as EEG Conformer. The adaptation framework comprises pre-training the model on one dataset and then fine-tuning the model on another dataset. The performance evaluation on these two datasets exhibits the effectiveness and robustness of the proposed approach in the domain of EEG signal classification. In the future, the proposed method can be optimized to make it more parameter-efficient while increasing performance.

\bibliographystyle{splncs04}
\bibliography{paper.bib}

\end{document}